\documentclass[letterpaper,prl,twocolumn,showpacs,superscriptaddress]{revtex4}
\usepackage{graphicx}

\begin{document}

\title{Squeezing in the audio gravitational wave detection band}

\author{Kirk McKenzie} \affiliation{Center for Gravitational Physics,
Department of Physics, Faculty of Science, The Australian National
University, ACT 0200, Australia}

\author{Nicolai Grosse} \affiliation{Center for Gravitational Physics,
Department of Physics, Faculty of Science, The Australian National
University, ACT 0200, Australia} \affiliation{Quantum Optics Group,
Department of Physics, Faculty of Science, The Australian National
University, ACT 0200, Australia} 

\author{Warwick P. Bowen} \affiliation{Quantum Optics Group,
Department of Physics, Faculty of Science, The Australian National
University, ACT 0200, Australia}

\author{Stanley E. Whitcomb} \affiliation{LIGO Laboratory, Californian
Institute of Technology, Pasadena, California, 91125, USA}

\author{Malcolm B. Gray} \affiliation{Center for Gravitational
Physics, Department of Physics, Faculty of Science, The Australian
National University, ACT 0200, Australia}

\author{David E. McClelland} \affiliation{Center for Gravitational
Physics, Department of Physics, Faculty of Science, The Australian
National University, ACT 0200, Australia}

\author{Ping Koy Lam} \affiliation{Center for Gravitational Physics,
Department of Physics, Faculty of Science, The Australian National
University, ACT 0200, Australia} \affiliation{Quantum Optics Group,
Department of Physics, Faculty of Science, The Australian National
University, ACT 0200, Australia}

\begin{abstract}
We demonstrate the generation of broad-band continuous-wave optical
squeezing down to 200Hz using a below threshold optical parametric oscillator (OPO).  The squeezed state phase was controlled using a noise locking technique. We show that low frequency noise sources, such as seed noise, pump noise and detuning fluctuations, present in optical parametric amplifiers have negligible effect on
squeezing produced by a below threshold OPO. This low frequency squeezing is ideal for improving the sensitivity of audio frequency measuring devices such as gravitational wave detectors.
 \end{abstract}

\date{\today} \maketitle

Squeezed light was proposed for quantum noise reduction in interferometric gravitational wave (GW) detection over two decades ago\cite{Caves}. Since then, the first generation of long baseline GW detectors - LIGO\cite{Abbott}, VIRGO\cite{Caron}, GEO~600\cite{Luck} and TAMA 300\cite{Ando} have been built and recently begun operation. The second generation of detectors, such as Advanced LIGO\cite{Schumaker}, are currently in the late planning stages. The prediction that they will be quantum noise limited (QNL) across most of the GW signal band (10Hz-10$^4$Hz)   has led to further theoretical investigations into the use of squeezing \cite{Kimble2, Harms, Mavalvala} and other optical methods \cite{Braginsky, Buonanno1, Courty} for quantum noise reduction.
However, to date only one experimental demonstration of quantum noise reduction in a GW detector configuration has been reported \cite{McKenzie}, and that result was obtained well above the GW signal band.

To be applicable to GW detectors, the requirements on squeezing include; continuous-wave (CW) at 1064nm, squeezed at the GW signal frequency, compatible with readout techniques \cite{Buonanno}, controllable phase and a high level of squeezing ($\sim$10dB).  Although squeezed light was first demonstrated in 1985 \cite{Slusher} a CW squeezed source at audio frequencies has not been reported until now. Laser relaxation oscillation and other technical noise sources have typically confined squeezing to the MHz range. 

Two of the most successful systems for squeezing generation have been the optical parametric oscillator (OPO) and optical parametric amplifier (OPA), for example see \cite{Lam, Xiao}. OPA and OPO have the same underlying second order
nonlinearity, however, they differ in that the OPA process has a coherent
seed field at the fundamental wavelength, whereas the OPO does not and
is seeded only by vacuum fluctuations. In theory OPO/A systems can produce squeezed states that fulfill the GW detector requirements outlined above. Experiments to date have been able demonstrate each requirement - except squeezing in the GW signal band. The lowest frequency CW squeezing experiments reported so far include Bowen {\it et al} \cite{Bowen}, Schnabel {\it et al} \cite{Schnabel} and Laurat {\it et al} \cite{laurat1} demonstrating squeezing down to 220kHz, 80kHz and 50kHz, respectively.  These experiments used either OPO or OPA, with \cite{Bowen,Schnabel} relying on common mode noise cancellation techniques.

In this letter we report the generation of high purity squeezing by a below threshold OPO at sideband frequencies down to 200Hz, continuous from 280Hz to well above 100kHz, covering a large portion of the audio GW detection band.  The phase of the squeezed vacuum relative to the homodyne detector was controlled without a carrier by using a noise dither locking technique, see for example \cite{laurat1}.  The inferred squeezing level (adjusted for detection efficiency) at the OPO output was 5.5dB$\pm$0.6dB below the shot noise limit (SNL) with inferred purity of 1.3$\pm$0.1, close to a minimum uncertainty state.  We compare OPO and OPA operation and find that the presence of a coherent seed field leads to dramatic degradation of the squeezing at low frequency, due to technical noise coupling.  The system operating as an OPO displays immunity to the same technical noise that degrades OPA squeezing.

Amplitude (+) and phase (-) quadrature variances, $V_{\rm sqz}^\pm
= \langle (X^{\pm}_{sqz})^{2}\rangle$, of the singly resonant OPO/A on resonance can
be modeled using linearized formalism by;
\begin{eqnarray}
V_{\rm sqz}^\pm (\omega) &=&  \bigg[ C_s V_{s}^\pm
(\omega)+ C_l V_l^{\pm}(\omega)  + C_v^\pm(\omega)
V_v^\pm(\omega)\nonumber \\ &+&\alpha^2 \big[C_p V_{p}^\pm (\omega)
+ C_\Delta^\pm V_\Delta(\omega) \big] \bigg] /
\big|D^\pm (\omega) \big|^2
\label{SQZEQN}  
\end{eqnarray}
where $V_{sqz}^{\pm}$ contains contributions from; the seed field, $V_s^\pm$; the pump field, $V_p^\pm$; vacuum fluctuations from intra-cavity loss, $V_l^\pm$; vacuum fluctuation entering through the output coupler, $V_v^\pm$; and noise due to detuning fluctuations in the cavity \cite{squash}, $V_\Delta$, which arise from such source acousto-mechanical disturbances.  Other sources, such as phase matching fluctuations, are not discussed here.  The denominator and coupling coefficients are given by \cite{Bowen_thesis} \footnote{ The notation $[{3 \atop 1}]$  means multiply by 3 for the amplitude quadrature or 1 for the phase quadrature.};

\begin{eqnarray}
D^{\pm}(\omega) &=& i\omega+\kappa_a+\Big[{{\small 3} \atop
{\small 1}}\Big]\epsilon^2 \alpha^2/{(2 \kappa_b)} \mp \epsilon \beta \nonumber \\ 
C_s&=& 4 \kappa_{in}^a \kappa_{out}^a \nonumber \\ \nonumber C_l&=&4\kappa^a_l\kappa_{out}^a \\ \nonumber
C_v^{\pm}(\omega)& =&  |2\kappa_{out}^a-D^\pm(\omega)|^2\\ \nonumber
C_p&=& 4 \kappa_{out}^{a} \kappa_{in}^b (\epsilon/\kappa_b)^2\\ \nonumber
C_\Delta^\pm& =&8\kappa_{out}^a\Big[{{\small 0} \atop{\small 1}}\Big]\nonumber
\end{eqnarray}

\begin{figure}[t!]
\begin{center}
\includegraphics[width=8.6cm]{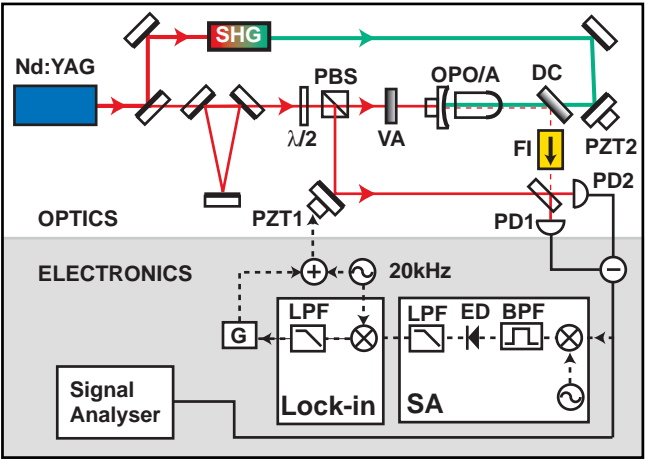}
\end{center}
\caption{ Schematic of the experiment.  The experiment was operated in
both OPO and OPA modes.  The OPA seed power was varied using a
variable attenuator (VA).  The OPO cavity was isolated from
backscatter off the photodetectors using a Faraday isolator (FI).  The
control electronics for the homodyne detection phase are indicated by
dashed lines.  SA-Spectrum analyzer, BPF-Band pass filter, ED-Envelope detector, LPF-Low pass filter,
G-Gain stage, SHG-second harmonic generator, OPO/A-optical parametric
oscillator/amplifier, PZT-piezo electric transducer, PBS-polarizing
beamsplitter, $\lambda$/2-half-wave plate, DC-dichroic mirror, PD-Photodetector. }
\label{EXP}
\end{figure}

where the intra-cavity fundamental field, $\alpha=\sqrt{n}$, where $n$
is the mean intra-cavity photon number. The parameters $\kappa_{out}^a,
\kappa_{in}^a$ and $\kappa_l^a$ are the decay rates of $\alpha$ due to the
output coupler, input coupler and loss, respectively. $\kappa_{in}^b$ is the decay rate of the intra-cavity second harmonic field, $\beta$, due to the input coupler. $\kappa_a,
\kappa_b$ are the total decay rates for $\alpha$ and $\beta$.  $\epsilon$ is the non-linear coupling parameter and $\omega$ is a small frequency shift relative to the
carrier frequency. The first three terms in Eq. \ref{SQZEQN} are standard contributions from the seed noise entering through the input coupler, vacuum fluctuations due to intra-cavity loss and vacuum fluctuations entering through the output coupler. The last two terms scale with $\alpha^2$, and show an important difference between OPO and OPA operation. That is; the fluctuations of the pump, $V^\pm_p$, and detuning, $V_\Delta$,  are coupled into the squeezed field, $V_{sqz}^\pm$ via the beat with the intra-cavity fundamental field, $\alpha$.  Thus, a below threshold OPO ($\alpha = 0$) should immune to these two noise sources to first order.  

\begin{figure}[t!]
\begin{center}
\includegraphics[width=8.6cm]{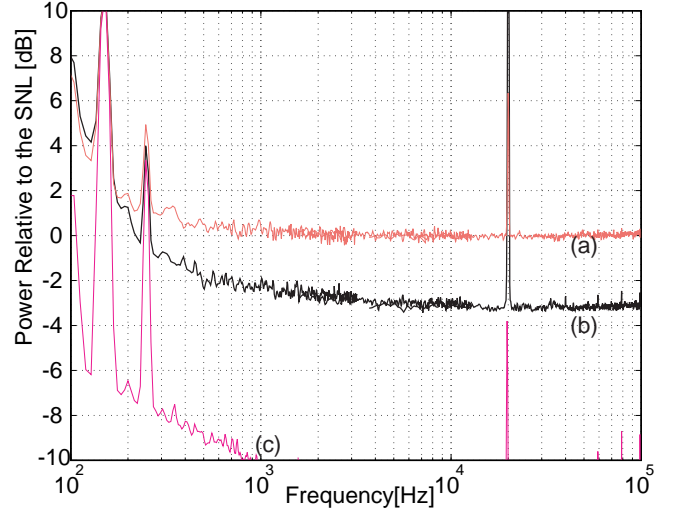}
\end{center}
\caption{Measured noise spectra for (a) the quantum noise limit, (b)
the squeezed light and (c) the electronic noise of the homodyne
detection system.The traces are pieced together from
three FFT frequency windows: 100Hz-3.2kHz, 1.6kHz-12.8kHz, and 3.8kHz-100kHz.  Each point is the averaged RMS value of 500, 1000 and 2000 measurements made in the respective ranges.  The RBW of the three windows was 8Hz, 32Hz and 128Hz,
respectively.  The electronic noise was -12dB below the quantum
noise from 10kHz-100kHz.  The 20kHz peak arises from the homodyne
locking signal.  Peaks at 50Hz harmonics are due to electrical mains supply.} \label{OPO}
\end{figure}

A schematic of the experiment setup is in Fig.~\ref{EXP}.  A Nd:YAG
laser operating at 1064nm was split into two beams.  One beam is used
to pump the second harmonic generator (SHG).  The other was spatially
and temporally filtered by a mode-cleaner cavity\cite{Rudiger} and used as
the seed beam for the OPA and as local oscillator (LO) for the
homodyne detector.  The OPO/A and SHG are constructed out of type-I
phase-matched MgO:LiNbO$_{3}$ hemilithic crystals.  The curved surface
of these crystals were coated for high reflectivity (HR) and the flat
surface coated for anti-reflectivity (AR) at both 532nm and 1064nm.
In both the SHG and OPO/A, standing-wave cavities were formed at 1064nm between the HR surface of
the crystal together with an external mirror of reflectivity  $R_{\rm IR} = 96$\% at 1064nm and $R_{\rm GR} < 4$\% at 532nm.  The OPO/A was
pumped with 100mW of 532nm light which double passed through the crystal
giving a measured classical gain of 5.  Results were taken in OPO mode,
and in OPA mode whilst varying the seed power from 1nW-6$\mu$W.  The squeezed state was detected on the homodyne detection system which had 96.5\% fringe visibility.  Whilst in OPO operation a Faraday isolator was inserted between the OPO
cavity and the photodetectors to reduce LO backscattered light.  The
photodetectors were built around ETX 500 photodiodes with 93\% quantum
efficiency.  The common mode rejection of the homodyne detector was over 55dB. 

\begin{figure}[t!]
\begin{center}
\includegraphics[width=8.6cm]{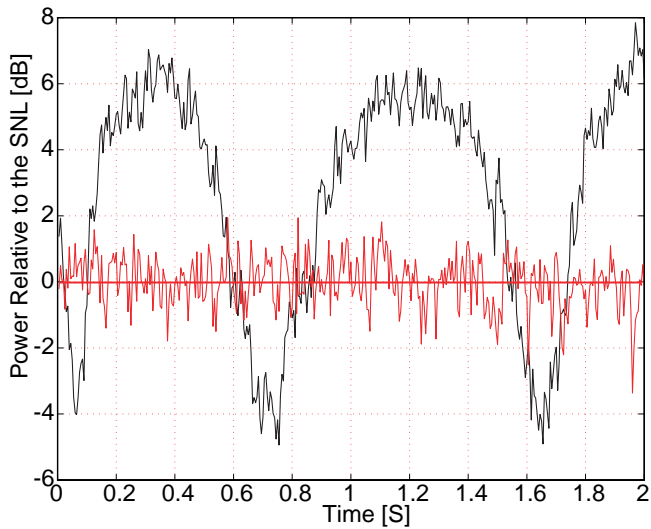}
\end{center}
\caption{The squeezed state at 11.2kHz as the phase of the homodyne is
varied.  RBW=1kHz, VBW=30Hz.  Electronic noise (9dB below SNL) was subtracted from
the data.} \label{OPO_2}
\end{figure}

The control electronics for the homodyne detection phase are indicated by dashed lines in Fig.~\ref{EXP}. This error signal was generated by dithering the LO phase and demodulating the difference photocurrent noise power. The noise power was detected using a spectrum analyzer (Agilent-E4407B, zero span at 2MHz, RBW=300kHz, VBW=30kHz) then demodulated with a lock-in amplifier (Stanford Research Systems (SRS)-SR830) and filtered before being fed back to PZT1. The stability of the homodyne detection phase enabled us to take results without locking the OPO/A cavity - which typically stayed on resonance for 10
seconds.  All data, except that in Fig.~\ref{OPO_2}, was recorded on a dynamic signal analyzer (SRS-SR785).

The OPO squeezing spectrum from 100Hz-100kHz is shown in
Fig.~\ref{OPO} \footnote{The OPO squeezing spectrum continues to 10's of MHz, as reported in \cite{Bowen_thesis}.}.   Trace (a) shows the quantum noise limit of the homodyne detection system.  The measurement of the squeezed light is shown in Trace (b). Trace (c) shows the electronics noise of the detection system.  The roll-up in noise power in traces (a) and (b) is attributable to the imperfect homodyne cancellation at low frequencies. We see that broad-band squeezing is obtained from 280Hz to 100kHz, with the exception of a locking signal peak at 20kHz.  Squeezing could not be measured at 150Hz and 250Hz due to power supply harmonics in the electronic noise.
More than 0.5dB of squeezing is still observable at frequencies around 200Hz.

\begin{figure}[t!]
\begin{center}
\includegraphics[width=8.6cm]{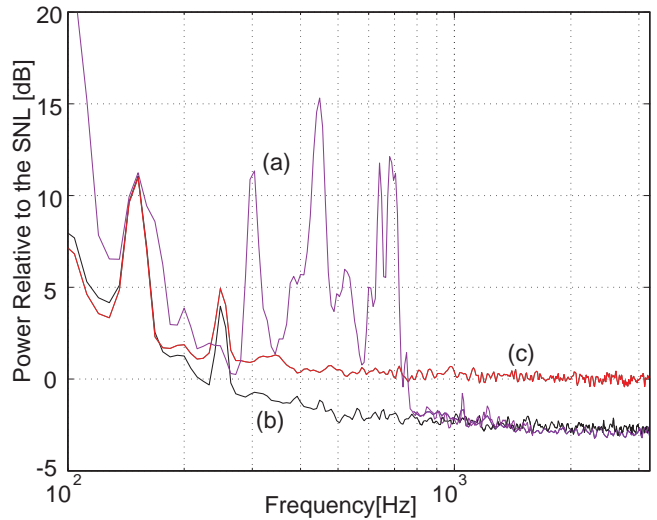}
\end{center}
\caption{The OPO spectrum 100Hz-3.2kHz without (a) and with
(b) the Faraday isolator between the OPO cavity and homodyne
detector.  RBW=8Hz, Number of RMS averages for (a):400
for (b) and (c):500.  Electronic noise (not shown) was not subtracted.} \label{OPA_2}
\end{figure}

Figure~\ref{OPO_2} shows the OPO squeezed state at 11.2kHz as the
homodyne phase was varied.  The measured squeezed state purity is
$V^+V^-=1.6$$\pm$0.2.  The squeezing and purity at the output of
the OPO can be inferred by taking into account the photodetection and homodyne efficiencies and optics loss.  The inferred purity is $V^+V^-=1.3$$\pm$0.1 and the inferred squeezing is $V_{sqz}=-5.5$dB$\pm$0.6dB.

Figure~\ref{OPA_2} shows a more detailed analysis of the squeezing
spectrum at the lowest frequency window of 100Hz-3.2kHz.  Trace
(a) shows the squeezing spectrum obtained without an isolator
in front of our homodyne detection system.  We observed large peaks
between 300Hz and 700Hz due to low frequency noise contamination.
This contamination is attributed to light from the LO  backscattered from the
photodetectors feeding into the OPO cavity.  We note that even
with the photo-detectors tilted away from retro-reflection, the
scattering from the front face of the detectors, which is estimated to
be of the order of ~1pW, is sufficient to seed the crystal and causes
parametric amplification.  With the Faraday isolator in place noise coupling via parametric amplification is eliminated, as shown by Trace (b). The
squeezed beam experiences an extra 9\% transmission loss through the isolator.  Similar to Fig.~\ref{OPO}, 
electronic noise is still present at 150Hz and 250Hz.

\begin{figure}[t!]
\begin{center}
\includegraphics[width=8.6cm]{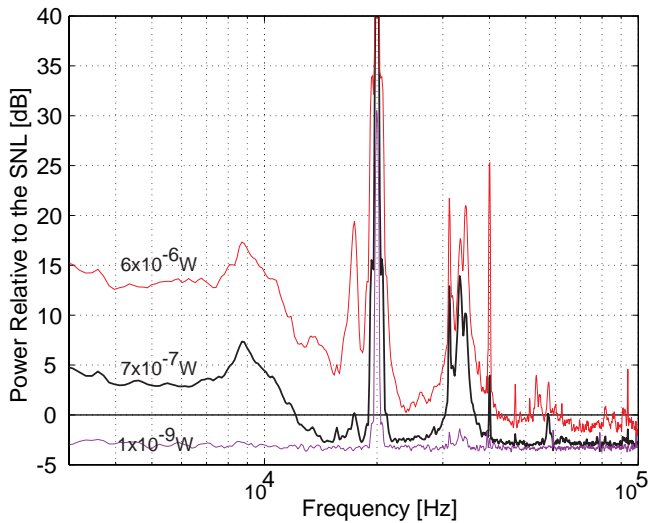}
\end{center}
\caption{The OPA spectrum 2kHz-100kHz with different OPA seed powers. RBW=128Hz, Number of RMS averages:1000, except for 6$\mu$W seed
power which had 500 RMS averages.  Electronic noise 
(at -12dB) was subtracted from all traces.}
\label{OPA}
\end{figure}

\begin{figure}[t!]
\begin{center}
\includegraphics[width=8.6cm]{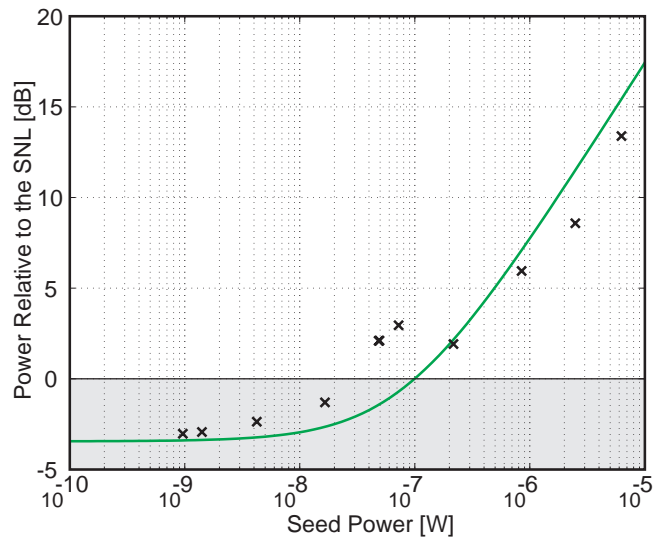}
\end{center}
\caption{The average noise power from 5-6kHz as a function of seed power, experimental data indicated by `x', model fit given by line. Electronic noise (at -12dB) was subtracted from all data.}
\label{OPA_3}
\end{figure}

The OPA spectrum from 2kHz-100kHz  is shown in Fig.~\ref{OPA} for three different seed powers, 1nW, 700nW and  6$\mu$W.  The data were recorded for optimal squeezing at 50kHz. The 1nW seed power spectrum resembles the OPO spectrum, with the exception of one added feature at 34kHz. The spectrum of the 700nW seed power shows the feature at 34kHz has increased in amplitude with additional excess noise at other frequencies thereby limiting squeezing to above 10kHz.   The feature at 8kHz was also present the pump intensity noise spectrum and is expected to have coupled into the squeezed field via the intra-cavity fundamental field. As the seed power was increased further, the noise floor and features in the spectrum continued to increase; by seed power 6$\mu$W there is no longer squeezing below 40kHz.  The noise power increase of the OPA spectrum with seed power is evident in Fig.~\ref{OPA_3}, which shows the mean noise power between 5-6kHz as a function of seed power.  The experimental points indicated by `x' can be compared with a model which has linear dependence on seed power, given by the solid line. Although there are large uncertainties expected in the experimental data, since the OPA cavity was not locked, the data is not inconsistent with the linear trend predicted by the theory in Eq.~\ref{SQZEQN}.  

In summary, we have presented results demonstrating OPO squeezed vacuum down to 200Hz. The phase of squeezed vacuum was controlled using a noise locking technique. Such squeezed states could already be of use for reduction of shot noise in first generation GW detectors. The comparison of the OPO and OPA results highlights the immunity of OPO and the sensitivity of OPA to technical noise. The direction of our future research will be to implement an OPO cavity lock, increase the level of squeezing and probe frequencies lower than 100Hz for use in second generation GW detectors. 

In addition to the possible application to GW detectors, low frequency squeezed light could potentially be used to enhance many measurement devices with optical readout.  These include 
atomic force microscopes\cite{Treps} and thermo-optical spectrometers\cite{Boccara}. Many long-standing experimental goals in quantum optics, such as the inhibition of atomic decay\cite{Gardiner} and sub-Doppler cooling of two-level atoms\cite{Shevy}, could also be facilitated using broadband low frequency squeezing.

We thank Keisuke Goda, Julien Laurat and Nicolas Treps for useful 
discussions. This research was supported by the Australian Research Council. SW was supported by the United States National
Science Foundation under Cooperative Agreement No.
PHY-0107417. This paper has been assigned LIGO
Laboratory document number LIGO-P040018-00-R.

\end{document}